\documentclass[a4paper]{jpconf}
\usepackage{epsfig}
\usepackage{graphicx}
\usepackage{dcolumn}
\usepackage{bm}
\newcommand{\beq}{\begin{equation}}
\newcommand{\eeq}{\end{equation}}
\newcommand{\be}{\begin{eqnarray}}
\newcommand{\ee}{\end{eqnarray}}
\newcommand{\WmunuA}{W^{\mu\nu}}
\newcommand{\Lmunu}{L_{\mu\nu}}

\newcommand{\bbox}[1]{%
{\mbox{\boldmath {$#1$}}}
}
\def\g{g_{ij}}

\def\met{\frac{1}{2}}

\def\bi{\bibitem}

\def\qm{{\bf q}}

\newcommand{\bea}{\begin{eqnarray}}
\newcommand{\eea}{\end{eqnarray}}

\begin{document}
\title{Weak Response of Nuclear Matter at low Momentum transfer}
\author{Nicola Farina}
\address
{INFN, Sezione di Roma. I-00185 Roma, Italy\\}
\date{\today}
\ead{nicola.farina@roma1.infn.it}
\begin{abstract}
A quantitative understanding of the weak nuclear response is a prerequisite for the computer simulations of astrophysical phenomena like supernov$\ae$ explosions and neutron star cooling. In order to reduce the systematic uncertainties associated with the simulations, a consistent framework, able to take into account dynamical correlation effects, is needed to compute neutrino-nucleon and neutrino-nucleus reaction rates. 
In this paper we describe the many-body theory of the weak nuclear response at low energy regime. We show how to include both short and long correlations effects in a consistent fashion.  
\end{abstract}
\section {Introduction}
The description of neutrino interactions at low momentum transfer ($E_\nu \sim$ 10 MeV) with nuclei, and nuclear matter in general, is relevant to the study of many different problems, from supernov$\ae$ explosions\cite{fryer} to neutron star cooling \cite{yako}.

The systematic uncertainty associated with computer simulations depends heavily on the values of the neutrino-nucleon and neutrino-nucleus reaction rates used as inputs.

Nuclear many body theory provides a scheme allowing for a consistent treatment of 
neutrino-nucleus interactions. Within this approach, nuclear 
dynamics is described by a phenomenological hamiltonian, whose structure is completely determined 
by the available data on two- and three-nucleon systems, and both short and long range dynamical correlations are taken into account. 

Over the past decade, the formalism based on correlated wave functions, originally proposed to 
describe quantum liquids \cite{feenberg}, has been employed to carry out highly accurate 
calculations of the binding energies of both nuclei and nuclear matter, using either 
the Monte Carlo method \cite{WP,GFMC,AFDMC1} or the cluster expansion formalism and 
the Fermi Hypernetted Chain integral equations \cite{APR1,gpc1,gpc2}.

A different approach, recently proposed in Refs. \cite{shannon2,valli} exploits the 
correlated wave functions to construct an effective interaction suitable for use
in standard perturbation theory. This scheme has been employed to obtain a variety 
of nuclear matter properties, including the neutrino mean free path \cite{shannon2} and the 
transport coefficients \cite{valli,transport}.

In this work we describe the application of the formalism based on correlated wave functions
and the effective interaction to the calculation of the weak response of uniform nuclear matter.

\section{Many-body theory of the weak nuclear response}
\label{nuA-formalism}

\subsection{Neutrino-nucleus cross section}
The differential cross section of the process
\beq
\nu_\ell + A \rightarrow \ell + X \ , 
\label{process:e}
\eeq
in which a neutrino carrying initial four-momentum $k\equiv(E_\nu,{\bf k})$ interacts with a nuclear target, producing a lepton in a state of four-momentum
$k^\prime\equiv(E_\ell,{\bf k}^\prime)$, the target final state 
being undetected, can be written in Born approximation as (see, e.g., 
Ref. \cite{PRD})
\beq
\frac{d\sigma}{d\Omega_\ell dE_\ell} = \frac{G^2}{32 \pi^2}\
\frac{|{\bf k}^\prime|}{|{\bf k}|}\
 L_{\mu \nu} W^{\mu \nu}\ ,
\label{nu:cross:section}
\eeq
where $G=G_F \cos \theta_C$, $G_F$ and $\theta_C$ being Fermi's coupling constant and
Cabibbo's angle. The leptonic tensor, that can be written, neglecting all lepton mass, as
\beq
\Lmunu = 8 \left[ k_\mu k_\nu^\prime + k_\nu k_\mu^\prime -g_{\mu\nu} (k k^\prime)-i\epsilon_{\mu\nu\alpha\beta} k^\alpha k^{\prime\beta} \right] \ ,
\eeq
is completely determined by lepton
kinematics, whereas the nuclear tensor $\WmunuA$ contains all the information 
on target structure. Its definition involves the initial
and final hadronic states $ |0\rangle$ and $|X \rangle$, carrying 
four-momenta $p_0$ and $p_X$, respectively, as well as the nuclear 
weak current operator $J^\mu$:
\beq
\WmunuA = \sum_X \langle 0 | J^\mu | X\rangle 
 \langle X | J^\nu | 0\rangle \delta^{(4)}(p_0+q-p_X) \ ,
\label{e:hadrten}
\eeq
where the sum includes all hadronic final states.

In the low momentum transfer regime, we expect the non relativistic approximation to be applicable. Within this approach, the initial and final states can be obtained from nuclear many-body theory (NMBT), while the weak current entering the definition of the hadronic tensor [\ref{e:hadrten}] is expanded in powers of $|{\bf q}|/m$. At leading order, the resulting response can 
be written in the simple form 
\beq
S(\qm,\omega)=\frac{1}{N}\sum_n\langle 0|O_{\qm}^\dagger|n\rangle\langle
n|O_{\qm}|0\rangle\delta(\omega+E_0-E_n) \ .
\label{ris:def}
\eeq
where, in the case of charged current interactions, $O_{\qm}$ is the operator corresponding to 
Fermi or Gamow-Teller transitions, whose expressions in the coordinate space are:

\bea
\nonumber
O^F_i(\qm)= g_V \delta( {\bf r}_i - {\bf r}_i^\prime ) \ 
 \ {\rm e}^{i\qm{\bf r}_i} \tau_i^+ \\
{\bf O}^{GT}_i(\qm) = g_A \delta( {\bf r}_i - {\bf r}_i^\prime ) \ {\rm e}^{i\qm{\bf r}_i}
\bbox{\sigma}_i\tau_i^+ \ ,
\label{def:opfgt}  
\eea
where ${\bf r}_i$ specifies the position of the $i-$th particle.

\subsection{Description of the initial and final states}

Understanding the properties of matter at densities comparable to the central density of
atomic nuclei is made difficult by {\it both} the complexity of the interactions {\it and} the approximations implied in any theoretical description of quantum mechanical many-particle 
systems.

The main problem associated with the use of the nuclear potential models  in a many-body calculation lies in the strong repulsive core of the NN force, which cannot be handled within standard perturbation theory. 

Within NMBT, a nuclear system is seen as a collection of
point-like protons and neutrons whose dynamics are described by the hamiltonian
\beq
H = \sum_{i} t(i) + \sum_{j>i} v(ij) + \ldots \ ,
\label{NMBT:H}
\eeq
where $t(i)$ and $v(ij)$ denote the kinetic energy operator and the {\it bare} NN potential,
respectively, while the ellipses refer to the presence of additional many-body
interactions.

Carrying out perturbation theory in the basis provided by the eigenstates of 
the noninteracting system requires a renormalization of the NN potential. This is 
the foundation of the widely employed approach developed by Br\"uckner, Bethe and Goldstone,
in which $v(ij)$ is replaced by the well-behaved G-matrix, describing NN scattering in the 
nuclear medium (see, e.g. Ref.\cite{Marcello}). 
Alternatively, the many-body Schr\"odinger equation, with the hamiltonian 
of Eq.(\ref{NMBT:H}), can be solved using either the variational method or 
stochastic techniques. These approaches have been successfully applied to the study of 
both light nuclei \cite{WP} and uniform neutron and
nuclear matter \cite{APR1,GFMC,AFDMC1,AFDMC2}.

Our work has been carried out using the scheme of correlated basis function (CBF) theory in which nonperturbative effects due to the short-range repulsion are embodied in the basis functions.

\begin{figure}[hbt]
\vspace{0.1cm}
\begin{center}
{\epsfig{figure=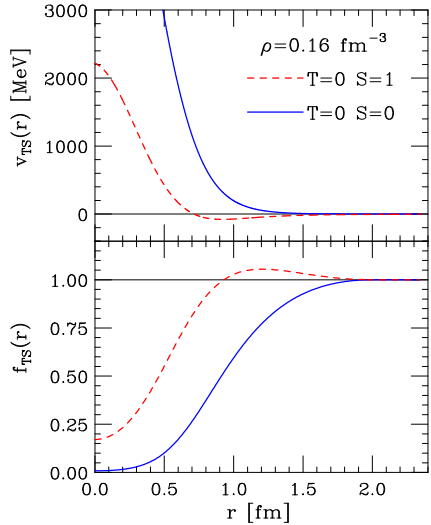,width=5.5cm }}
\vspace{0.1cm}
\caption{{
Interaction potentials (upper panel) and correlation functions (lower panel)
acting in the spin-isospin channels $S=0$ and $T=0$ (solid lines)
and $S=0$ and $T=1$ (dashed lines). The potential is the Argonne $v^\prime_8$ \cite{argonne} and
the correlation functions correspond to nuclear matter at equilibrium density. }
\label{corrf} }
\end{center}
\end{figure}

The {\em correlated} states of nuclear matter are obtained from the Fermi gas
(FG) states $\vert n_{FG}\rangle$
through the transformation \cite{feenberg,CBF1}
\begin{equation}
| n ) = \frac{ F |n_{{\rm FG}} \rangle }
{ \langle n_{{\rm FG}} | F^\dagger F | n_{{\rm FG}} \rangle^{1/2} } \ .
\label{def:corr_states}
\end{equation}
In the above equation, $|n_{{\rm FG}} \rangle$ is a determinant of single particle states 
describing $N$ noninteracting nucleons. The 
operator $F$, embodying the correlation structure induced by the NN
interaction, is written in the form
\begin{equation}
F(1,\ldots,N)=\mathcal{S}\prod_{j>i=1}^N f_{ij} \  ,
\label{def:F}
\end{equation}
where $\mathcal{S}$ is the symmetrization operator which takes care of the fact that, in general,
\beq
\left[ f_{ij} , f_{ik} \right] \neq 0 \ .
\eeq
The structure of the two-body {\em correlation functions} $f_{ij}$
must reflect the complexity of the NN potential. The shape of the radial functions $f^{n}(r_{ij})$ is determined through functional minimization of the expectation value of the nuclear hamiltonian in 
the correlated ground state
\beq
E^V_0 = ( 0 | H | 0 )  \ .
\label{def:E0V}
\eeq 
As an example, Fig.\ref{corrf} shows the radial dependence of the potentials ) and correlation functions 
acting in the spin-isospin channels $S=0$ and $T=0$ 
and $S=0$ and $T=1$. 

It has to be pointed out that the correlation operator of Eq.(\ref{def:F}) is defined such that, 
if any subset of the particles, say $i_1, \ldots i_p$, is removed far from the 
remaining $i_{p+1}, \ldots i_N$, it factorizes according to
\beq
F(1,\ldots,N) \rightarrow F_p(i_1, \ldots i_p) F_{N-p}(i_{p+1}, \ldots i_N) \ .
\eeq
The above property is the basis of the cluster expansion formalism, that allows 
one to write the matrix element of a many-body operator between correlated states as a sum, 
whose terms correspond to contributions arising from isolated subsystems 
({\it clusters}) involving an increasing number of particles.

\section{Correlation effects in the transition matrix elements}

Using correlated states implies severe difficulties in the explicit calculation of the weak 
matrix element. In the FG model, the nuclear response is non vanishing only when the final 
nuclear state differs from the initial state for the presence of a particle excited outside 
the Fermi sea and a hole in the Fermi sea. In the presence of correlations, which can 
induce virtual nucleon-nucleon scattering processes leading to excitation of nucleons to
states outside the Fermi sea, more complex scenarios must also be considered. 
For example, if the initial state has a two particle-two hole component, the final state can 
be a three particle-three hole state or, if the probe interacts with an 
excited nucleon, a two particle-two hole state.

In the following we will consider only the dominant transition, between the {\em correlated} 
ground state and a {\em correlated} one particle-one hole ($ph$) state. The corresponding 
weak matrix element can be written 
\beq
M_{ph}=\frac{\langle ph|F^\dagger O F|0\rangle}
{\langle ph|F^\dagger F|ph\rangle^\met\langle 0|F^\dagger F|0\rangle^\met} \ ,
\label{Mph}
\eeq
where $F$ is the correlation operator defined in Eq.(\ref{def:F}). Here the kets $|0\rangle$ 
and $|ph\rangle$ correspond to the ground and one particle-one hole Fermi Gas states, 
respectively, 
and $O=\sum_i O_i$, $O_i$ being the Fermi or the Gamow-Teller transition operator 
(see Eqs.(\ref{def:opfgt})).

Note the the $g_{ij}=f_{ij}-1$ is short ranged, and therefore its matrix elements are small. 
At two-body level, the cluster expansion of  Eq.(\ref{Mph}) yields \cite{phdtesi}
\beq
\langle ph |F^\dagger O F |0\rangle  \simeq  \langle ph|(1+\sum_{j>i}\g)O 
(1+\sum_{j>i}\g)|0\rangle \ .
\label{Nph}
\eeq
The above equation suggests the definition of an {\em effective operator} $ O^{eff}_{12}$, 
acting on Fermi Gas states. From 
\beq
\langle ph|F^\dagger O F|0\rangle=\langle ph|O^{eff}|0\rangle .
\label{oeff:def}
\eeq
it follows that, at the two-body cluster level (compare to Eq.(\ref{Nph}))
\beq
\frac{1}{N} O^{eff}_{12}=O_1 +\frac{N-1}{2}\{O_1+O_2,g_{12}\}+\frac{N-1}{2}[g_{12}(O_1+O_2)g_{12}] \ ,
\label{oeff:tb}
\eeq
where $\{A,B\}=AB+BA$ and $N$ denotes the number of particles.
Note that the $O^{eff}$ is a two-body operator, as it includes screening effects arising from 
nucleon-nucleon correlations. 

\begin{figure}[ht]
\centerline{\includegraphics[scale=0.45]{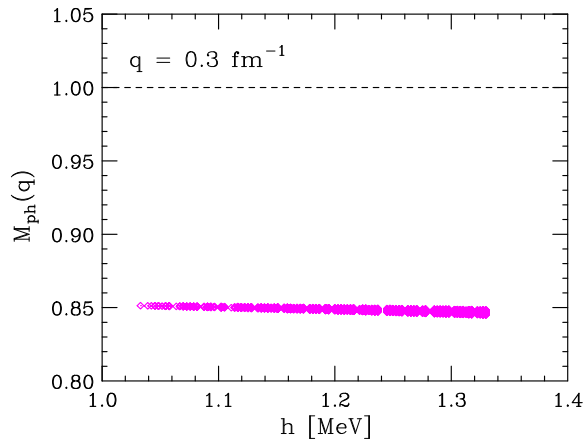}}
\caption{\small Fermi transition matrix element at $|{\bf q}|=0.3 \ {\rm fm}^{-1}$ 
as a function of the magnitude of hole momentum $|{\bf h}|$. The dashed horizontal 
line corresponds to the result of the FG model.  \label{quenching}}
\end{figure}

We have carried out the calculation of $M_{ph}$ on a cubic lattice, for a discrete set of $N_h$ states ${\bf h_i}$, satisfying the conditions ${\bf h_i}<k_f$ and ${\bf h_i}+{\bf q}>k_f $, $ k_f$ being the Fermi momentum. Figure \ref{quenching} shows a comparison between the correlated and the Fermi Gas matrix element for a Fermi transition. It has to be noticed that the effects of  correlations is enhanced in the calculation of the response, see Eq.(\ref{ris:def}), whose definition involves the square of the correlated matrix element.

\section{The effective interaction}
\label{sec:ei}

At lowest order of CBF, the effective interaction $V_{{\rm eff}}$ is {\em defined} by 
\beq
\label{veff:1}
\langle H \rangle = \langle 0_{FG} |F^\dagger H F|0_{FG}\rangle=\langle 0_{FG} | T_0 + V_{{\rm eff}}| 0_{FG} \rangle  \ .
\eeq
As the above equation suggests, the approach based on the effective interaction allows one 
to obtain any nuclear matter observables using perturbation theory in the 
FG basis. However, as discussed in the previous Section, the calculation of the hamiltonian 
expectation value in the correlated ground state, needed to extract $V_{{\rm eff}}$ from
Eq.(\ref{veff:1}), involves severe difficulties. In this work we follow the procedure developed in Refs. \cite{shannon1,shannon2}, whose 
authors derived the expectation value of the effective interaction by carrying out a 
cluster expansion of the rhs of Eq.(\ref{veff:1}), and 
keeping only the two-body cluster contribution. The resulting expression can be 
written
\begin{equation}
 \langle 0_{FG} | V_{{\rm eff}} | 0_{FG} \rangle = 
\sum_{i<j} \langle ij | v_{{\rm eff}}(12) | ij \rangle_a 
 = \sum_{i < j} \langle ij | f_{12} \left[ -\frac{1}{m} (\nabla^2 f_{12}) + 
v(12) f_{12} \right] | ij \rangle_a \ ,
\label{veff:2}
\eeq
where the laplacian  operates on the relative coordinate and the suffix $a$ denotes that the ket $|ij\rangle_a$ is anty-simmetryzed.

In the left panel of Fig.\ref{veff1} the central components  of the effective interaction obtained from the $v_8^\prime$ potential of Ref.\cite{argonne} at equilibrium density is compared to the corresponding component of the bare interaction. It clearly appears that screening effects due to NN correlations lead to a significant quenching.

\begin{figure}[h!]
\vspace{0.1cm}
\begin{center}
\psfig{file=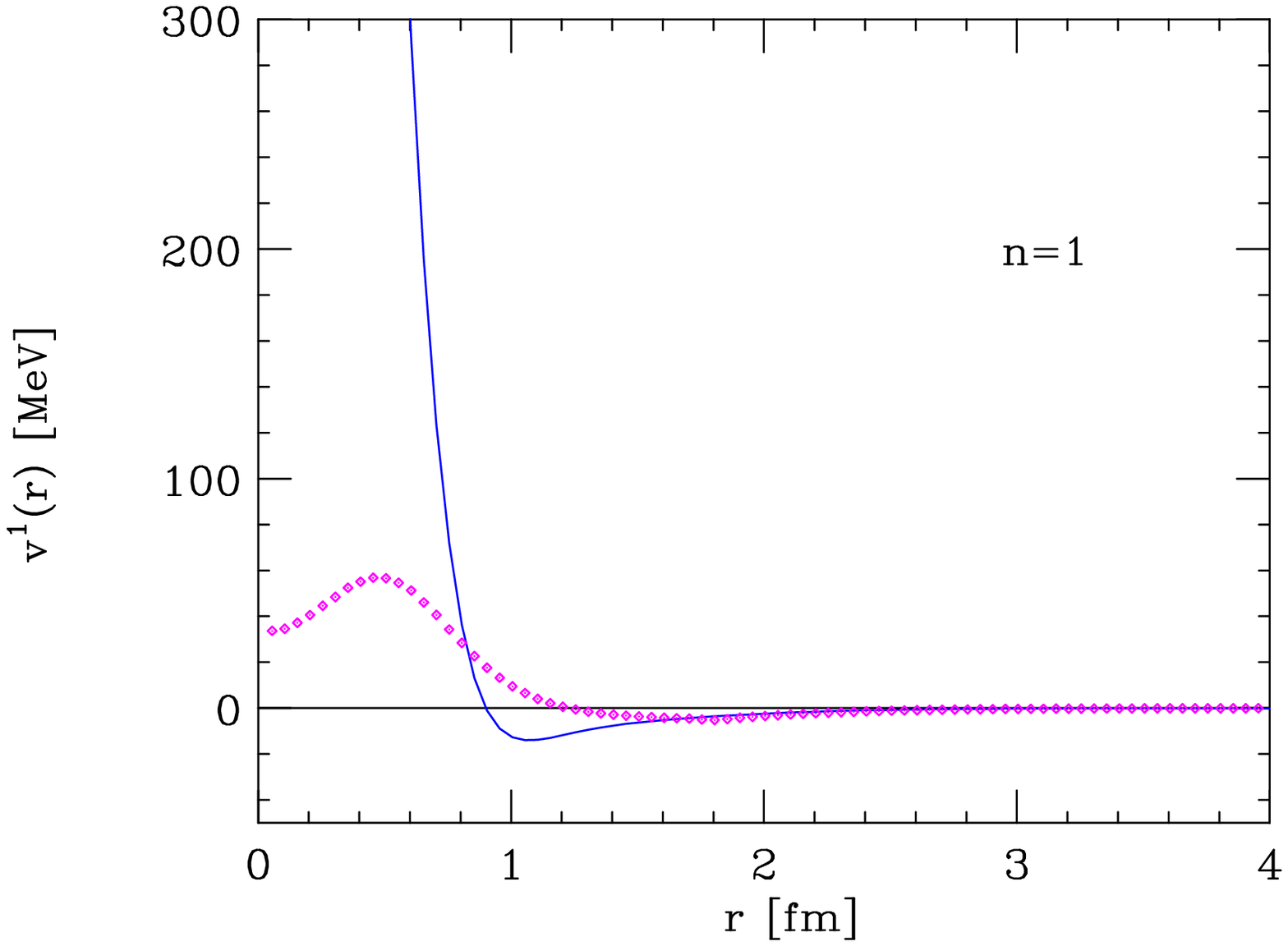,width=6.4cm,angle=0}
\includegraphics[scale=0.37]{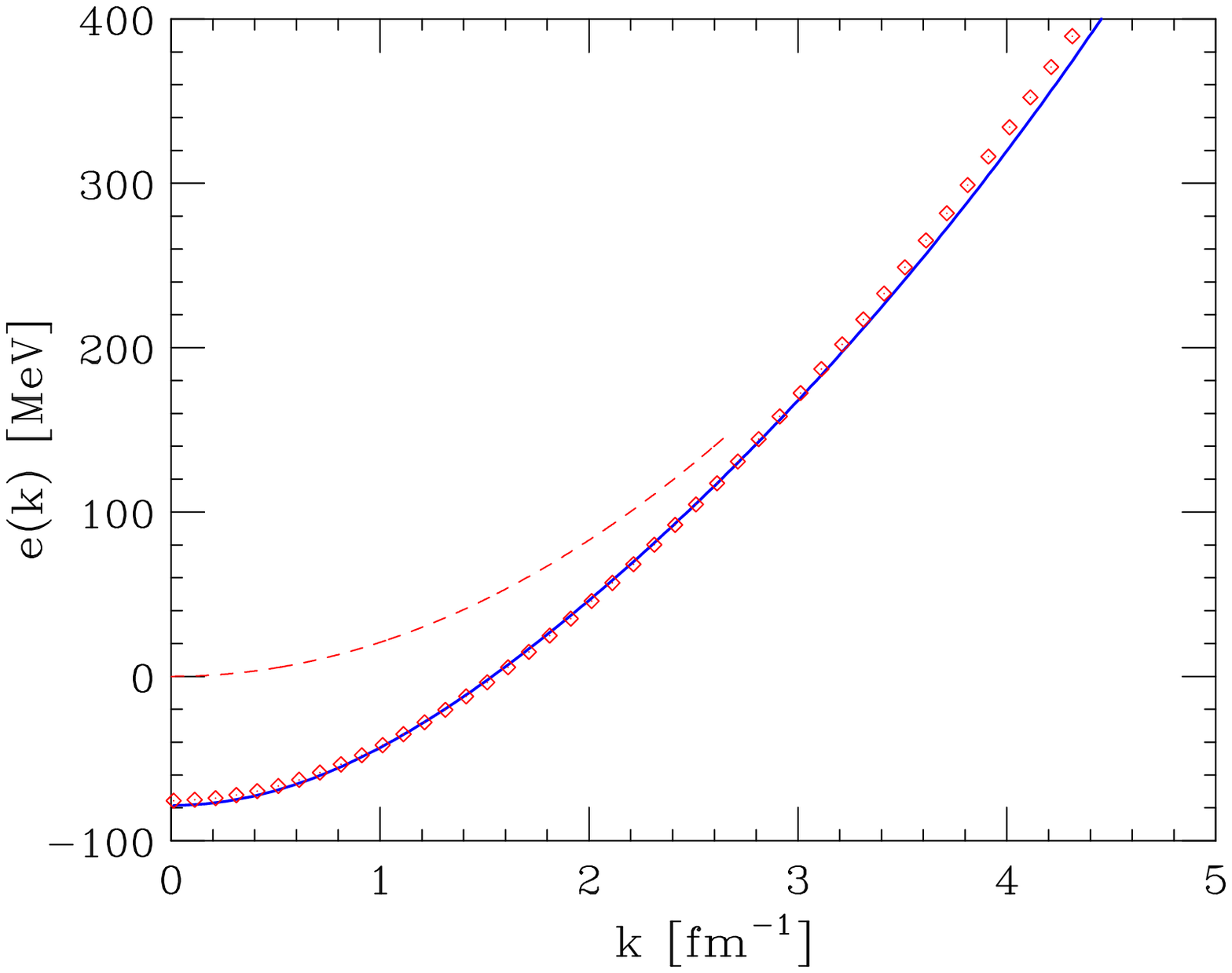}
\end{center}
\caption{{\small Left panel: Comparison between the central component of the bare Argonne $v^\prime_8$\cite{argonne} potential 
(dashed lines) and the effective potential defined by Eq.(\ref{veff:1}) (solid lines), 
calculated at nuclear matter equilibrium density. Right panel: momentum dependence of the single particle energies obtained from the CBF effective interaction in the Hartree-Fock approximation (solid line). The diamonds show the results of Ref.\cite{bob:ep}. For reference, the energies of the Fermi Gas model are also shown (red line):} \label{veff1} }
\end{figure}

Using the effective interaction described above, the single particle energies $e_{\bf k}$ in nuclear matter can be easily computed in  Hartree-Fock approximation. The risulting expression is:

\beq
e_{\bf k}=\frac{{\bf k}^2}{2m}+\sum_{h<k_f} \langle h k|V_{eff}|h k\rangle_a \ ,
\label{singlespectrum}
\eeq

The right panel of Fig. \ref{veff1} shows the energy spectrum for symmetric nuclear matter at equilibrium density. 

\section{Correlation effects on the response.}

\subsection{Short range correlations.}

We have calculated the nuclear response for a Fermi transition (Eqs.(\ref{ris:def}) and (\ref{def:opfgt})). The inclusion of interaction leads to sizable modifications of the FG response.
Correlation effects in the transition matrix elements, taken into account through
the use of the effective operator, produce a quenching of $\sim$ 15 \%. This feature is apparent 
in Fig. \ref{risp:src}, where the FG response, represented by the green line, is compared to that obtained 
using the effective operator in the calculation of $M_{ph}$ and the FG  single particle spectrum, 
represented by the red line. An even larger modification is produced by interaction effects on the single particle energies. Replacing the FG single particle energies  with the HF energies (Eq.\ref{singlespectrum}) leads to a sizable broadening of $\omega$ region corresponding to non-vanishing response as shown by the 
black line in Fig. \ref{risp:src}.

\begin{figure}
\begin{center}
\includegraphics[scale=.30]{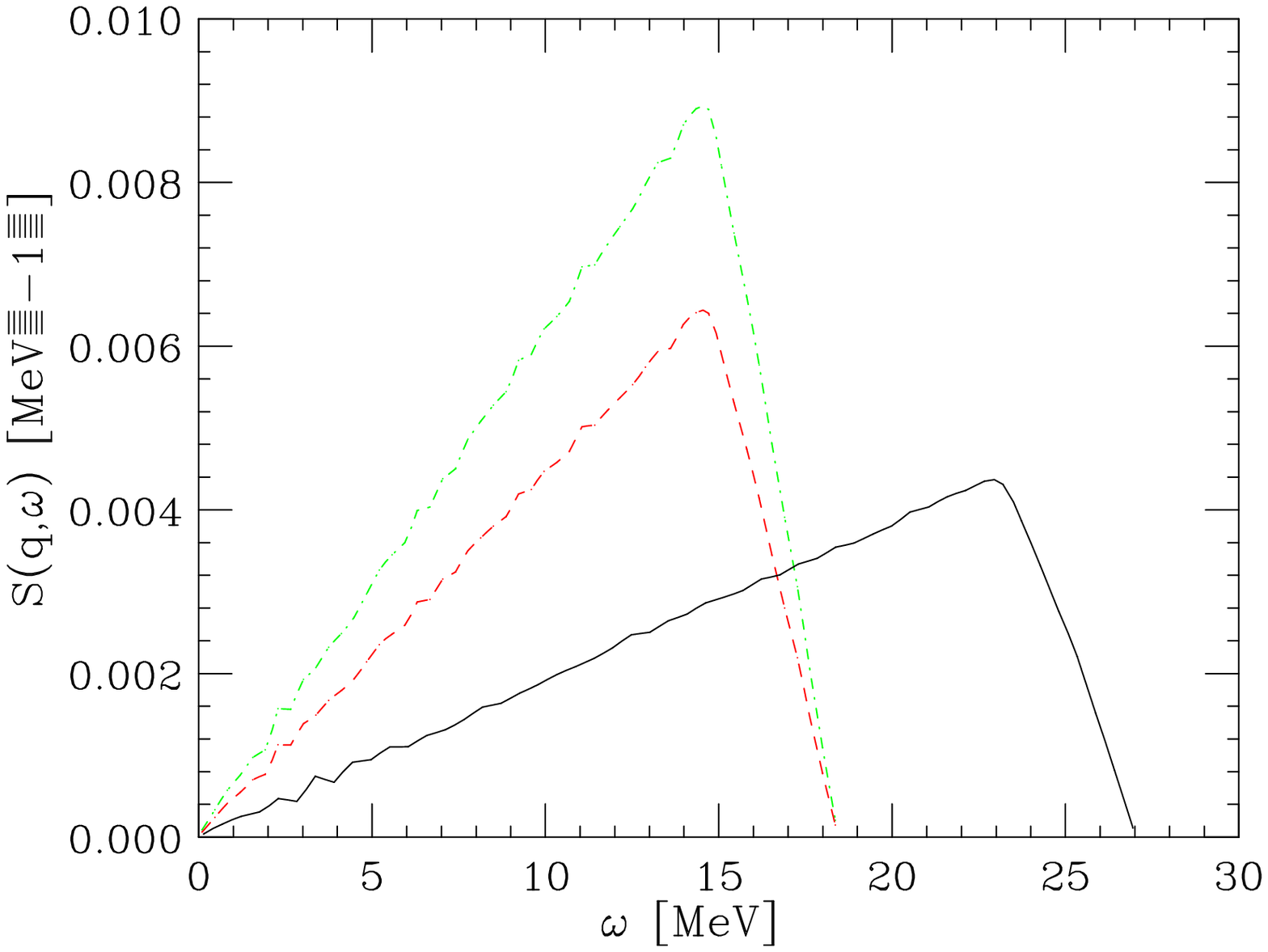}
\includegraphics[scale=.30]{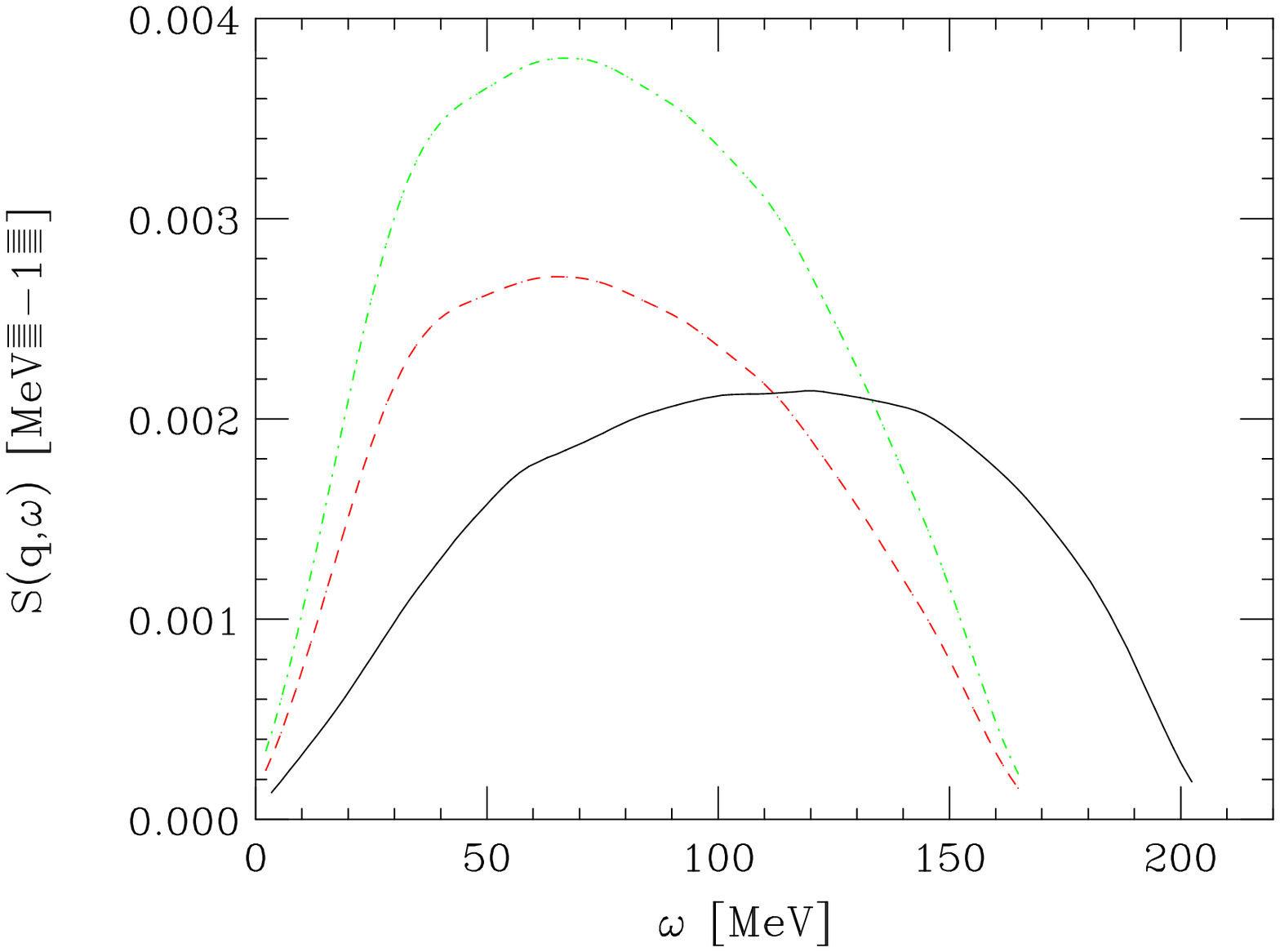}
\includegraphics[scale=.30]{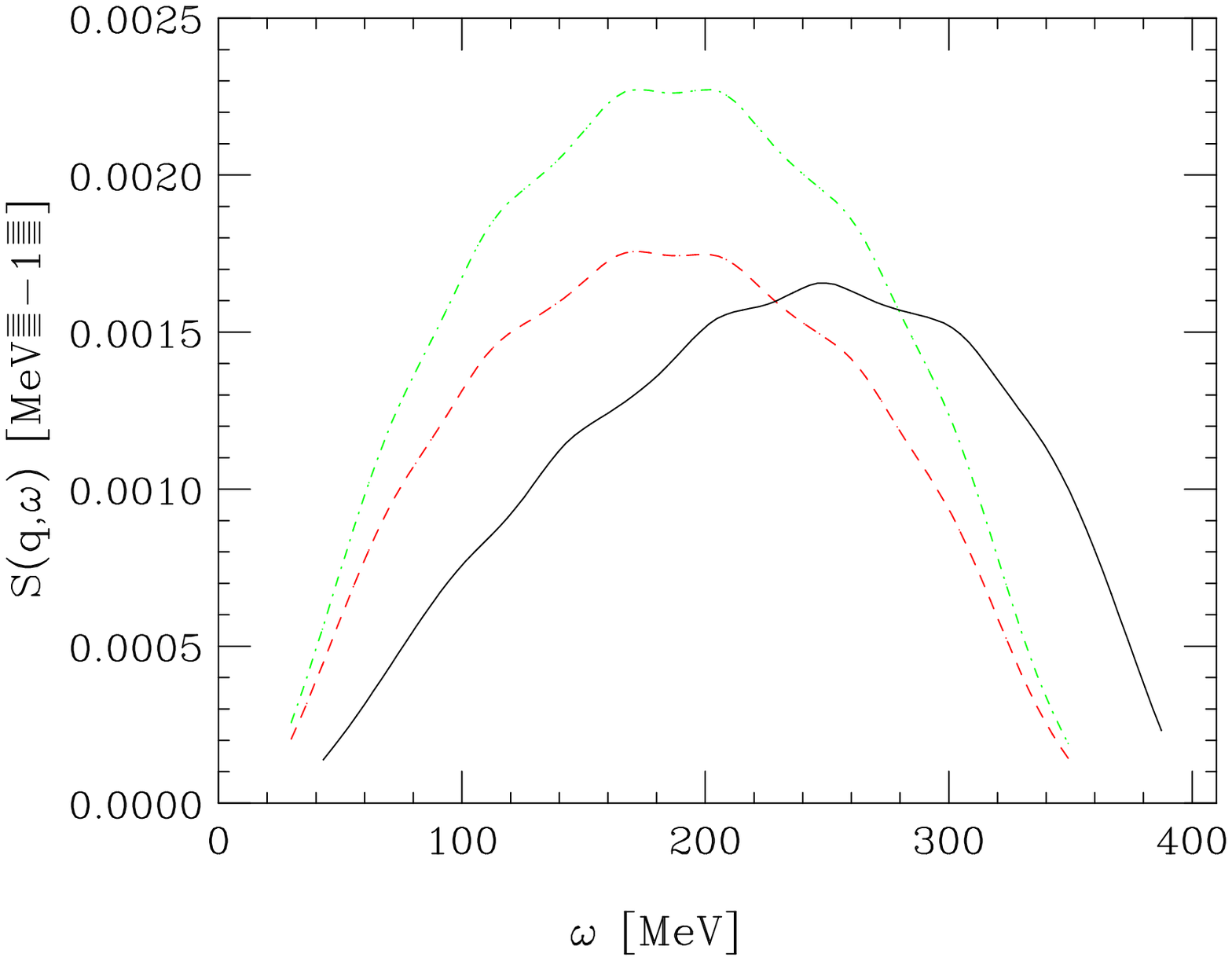}
\end{center}
\caption{Nuclear response for Fermi transition, for $\qm=.3$ fm$^{-1}$  (left), $\qm=1.8$ fm$^{-1}$ (center), and $\qm=3$ fm$^{-1}$ (right). Black line is full calculation. Confrontation with Fermi Gas (green line) and effective operator with Fermi Gas energy spectrum (red line) are shown.\label{risp:src}}
\end{figure}

\subsection{Long range correlations: the Tamm-Dancoff Approximation (TDA)}

In the previous section we have discussed the nuclear response in the correlated 
Hartree-Fock (HF) approximation, in which the bare Fermi transition operator is replaced 
by the effective operator of Eq.(\ref{oeff:def}) and the final state is assumed to
be a one particle-one hole state. 

It is important to realize that the FG one particle-one hole states, while being 
eigenstates of the HF hamiltonian, defined as 
\beq
H_{HF} = \sum_i e_i \ ,
\label{def:HHF}
\eeq
with $e_i$ given by Eq.(\ref{singlespectrum}), are not eigenstates of the full nuclear 
hamiltonian. As a consequence, there is a residual interaction $V_{res}$, which can be identified with the effective interaction defined in Section \ref{sec:ei}, that can 
induce transitions between different one particle-one hole states, as long as 
their total momentum {\bf q}, spin and isospin are conserved. 

In order to include the effects of these transitions, we use the TDA,
 which amounts to expanding the final state in the basis of
one particle-one hole states according to\cite{Boffi}
\beq
| f \rangle = | {\bf q}, \ T S M \rangle = \sum_{i} c^{TSM}_{i}
| {\bf h}_i ,\ {\bf p}_i = {\bf h}_i + {\bf q}, \ T S M \rangle \ ,
\label{tda:phexp}
\eeq
where $S$ and $T$ denote the total spin and isospin of the particle hole pair 
and $M$ is the spin projection. 

At fixed ${\bf q}$, the excitation energy of the state (\ref{tda:phexp}), 
$\omega^{f}$, as well as the coefficients $ c^{TSM}_{i}$, are obtained solving the eigenvalue equation 
\beq
H | f \rangle = (H_{HF} + V_{res})| f \rangle= \left(\sum_i e_i + \sum_{ij} v_{eff}(ij)\right)| f \rangle=(E_0 + \omega^{f}) | f \rangle \ ,
\label{eigenprob}
\eeq
where $E_0$ is the ground state energy. In the TDA, the response can be written as
\beq
S(\qm,\omega)=  \sum_{T,M,S}  \ \sum_{n} 
\left| \sum_{i}  (c^{TSM}_n)_i 
\langle {\bf h}_i, \ {\bf p}_i, \ T S M | O_{eff}(\qm)|0\rangle \right|^2  
\delta(\omega-\omega^{TSM}_n) \ ,
\label{tda:resp}
\eeq
where $(c^{TSM}_n)_i$ denotes the $i$-th component of the eigenvector belonging to the 
eigenvalue $\omega^{TSM}_n$.

We have performed the diagonalization of the Hamiltonian operator over a basis of $N_h\sim 3000$ one particle-one hole states for any spin-isospin channel. The appearance of an eigenvalue 
lying well outside the particle hole continuum, corresponding to a collective excitation, 
reminiscent to the plasmon mode of the electron gas, is clearly visible in Fig.\ref{res5}, where the calculation of the response for a Fermi transition within TDA is compared with the response of the correlated HF approximation. The  TDA response exhibits a sharp isolated peak corresponding to the 
collective mode, lying $\sim$ 4 MeV above the upper limit of the particle hole continuum.  

\begin{figure}[ht]
\begin{center}
\includegraphics[scale=0.37]{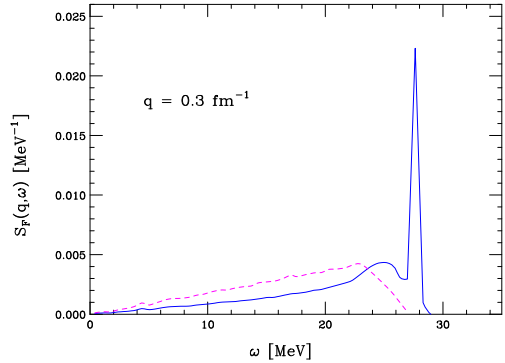}
\end{center}
\caption{\small Nuclear matter response in the TD approximation  at 
$|{\bf q}|=0.3 \ {\rm fm}^{-1}$ (solid line).  The dashed
lines show the results of the correlated HF approximation.
 \label{res5}}
\end{figure}

\section{Conclusions}
We have carried out calculations of the charged current weak response of nuclear matter in the low momentum transfer regime. The quantitative understanding of this quantity is required in many problems of physics,
ranging from simulations of supernov$\ae$ explosions to neutron star cooling.

The calculation has been performed using a many-body approach based on 
a realistic nuclear hamiltonian, including two- and three-nucleon interactions, 
yielding a good description of the properties of both the two-nucleon systems and
uniform nuclear matter. The response associated with Fermi transitions at low momentum transfer has been
calculated from an effective interaction, derived using the formalism of 
correlated basis functions and the cluster expansion technique. Our work improves 
upon existing effective interaction models 
in that it includes the effects of many-nucleon forces, which become sizable, 
at high density. The responses calculated within the correlated HF approximation show that the 
inclusion of short range correlations leads 
to a significant quenching of the transition matrix elements and shifts the strength
towards larger values of the energy transfer, $\omega$, for all values of $|{\bf q}|$.
At low momentum transfer, long range correlations have been taken into account 
within the TDA. The excitation of the coherent state can be clearly 
seen in our results at $|{\bf q}| = 0.3 \ {\rm fm}^{-1}$.

\end{document}